\documentclass[conference]{IEEEtran}
\IEEEoverridecommandlockouts

\usepackage{cite}
\usepackage{amsmath,amssymb,amsfonts}
\usepackage{algorithmic}
\usepackage{graphicx}
\usepackage{textcomp}
\usepackage{xcolor}
\usepackage{array}
\usepackage{booktabs}

\def\BibTeX{{\rm B\kern-.05em{\sc i\kern-.025em b}\kern-.08em
    T\kern-.1667em\lower.7ex\hbox{E}\kern-.125emX}}

\newcommand{\figmaybe}[2]{%
\IfFileExists{#1}{\includegraphics[width=\linewidth]{#1}}{%
\fbox{\begin{minipage}[c][2.0cm][c]{0.94\linewidth}
\centering\footnotesize #2\\Upload file: \texttt{#1}
\end{minipage}}}}

\begin{document}

\title{Common-Loss Parameter-Efficiency Analysis of MLP and KAN Neural Receivers for Digital Communications}

\author{
\IEEEauthorblockN{Sude Ertan, Osman Tokluoglu, Enver Çavuş}

\IEEEauthorblockA{
Elektrik-Elektronik Mühendisliği Bölümü,\\
Ankara Yıldırım Beyazıt Üniversitesi, Ankara, Türkiye\\
E-mail: suddertan@gmail.com, otokluoglu@aybu.edu.tr, ecavus@aybu.edu.tr
}
}

\IEEEpubid{
\makebox[\columnwidth]{%
\textbf{979-8-3195-4709-5/26/\$31.00 \copyright 2026 IEEE}%
\hfill}
\hspace{\columnsep}
\makebox[\columnwidth]{}
}
\maketitle

\begin{abstract}
Classical coherent binary phase-shift keying (BPSK) reception over additive white Gaussian noise (AWGN) channels is analytically well understood, and the optimum hard-decision detector is known. Therefore, the aim of using neural receivers in this study is not to replace the classical AWGN-BPSK detector. Instead, the AWGN-BPSK setting is deliberately selected as a theoretically verifiable benchmark for analyzing how compactly neural receiver architectures can represent a known decision behavior. This paper compares multi-layer perceptron (MLP) and Kolmogorov--Arnold Network (KAN) receivers using a multi-SNR Nyquist-rate BPSK dataset. The models are evaluated using bit error rate (BER), mean squared error (MSE), and trainable parameter count. Beyond reporting accuracy alone, this work emphasizes a common-loss parameter-efficiency perspective: when two receivers reach a comparable practical BER or loss region, the receiver with fewer parameters is more attractive for real-time deployment. The results show that both MLP and KAN receivers reproduce the expected AWGN-BPSK detection trend, while a compact KAN configuration reaches a comparable operating region with substantially fewer trainable parameters. In particular, the KAN receiver with 3281 parameters achieves a test BER of $2.45\times10^{-4}$, while the MLP baseline with 8513 parameters achieves a test BER of $2.50\times10^{-4}$. This corresponds to approximately 61.5\% fewer trainable parameters at a comparable BER operating point. This reduction is important for real-time neural receivers because it affects memory footprint, parameter access, inference latency, energy consumption, and hardware feasibility on embedded, software-defined radio, FPGA, ASIC, and edge communication platforms.
\end{abstract}


\section{Introduction}

Digital communication receivers are traditionally based on analytically derived signal processing methods. For coherent BPSK reception over an AWGN channel, the optimum detector is the sign decision rule, and the theoretical BER is available in closed form \cite{proakis2007digital,sklar2001digital}. Therefore, a neural network is not required to solve this specific detection problem. Instead, AWGN-BPSK is used here as a controlled and analytically verifiable benchmark, allowing receiver architectures, BER and MSE behavior, and parameter-efficiency trade-offs to be evaluated against a reliable reference with minimal ambiguity.

Learning-based physical-layer methods have been investigated for modulation recognition, channel estimation, equalization, demapping, end-to-end transceiver optimization, and receiver-side signal recovery \cite{oshea2017physical,dorner2018overair,honkala2021deeprx,simeone2018brief,wang2017wireless,oshea2018radio}. Although several deep learning-based detectors have also been proposed for Faster-than-Nyquist signaling with $\tau<1$ \cite{tokluoglu2025standalone,tokluoglu2026gru,tokluoglu2025domainaware,tokluoglu2026bigru}, this study intentionally considers conventional Nyquist-rate AWGN-BPSK. The objective is not to outperform FTN-oriented detectors, but to isolate the architectural parameter-efficiency question by comparing Kolmogorov--Arnold Networks and Multi-Layer Perceptrons using BER, MSE, and trainable parameter count.

MLPs are frequently used because they are simple and flexible universal approximators \cite{cybenko1989approximation}. However, dense MLPs may require many trainable parameters, which can limit their suitability for real-time and resource-constrained deployment. Recently, KANs have been introduced as an alternative neural architecture inspired by the Kolmogorov--Arnold representation theorem \cite{liu2025kan,kolmogorov1957representation,arnold1957functions}. Unlike MLPs, which apply fixed nonlinear activation functions at nodes, KANs employ learnable univariate functions on edges, typically represented using spline bases \cite{deboor1978splines}. This different parameter allocation mechanism motivates the investigation of KANs as compact neural receiver models.

The central claim of this paper is not that KAN universally outperforms MLP in terms of minimum error. Instead, KAN is evaluated as a parameter-efficient neural receiver architecture. In real-time communication systems, the model with the lowest offline MSE is not always the most practical receiver. A deployable receiver must satisfy target BER, latency, memory, power, and hardware constraints. Therefore, if two receivers reach a comparable practical BER region, the smaller model can be preferable.

The main contribution of this study is a controlled and implementation-aware comparison of MLP and KAN neural receivers for Nyquist-rate BPSK reception over AWGN channels. Within this scope, a theoretically verifiable BPSK receiver benchmark is constructed, and both receiver families are evaluated under identical data generation, preprocessing, training, validation, and test conditions. Instead of comparing the models only through offline regression accuracy, the study jointly considers BER, MSE, and trainable parameter count. This enables a common-loss based parameter-efficiency interpretation, where receivers that reach a comparable practical BER region are compared in terms of model compactness. The resulting analysis connects neural receiver performance with real-time deployment considerations such as memory footprint, parameter access, inference latency, energy consumption, and hardware feasibility on embedded, SDR, FPGA, ASIC, and edge communication platforms.

\section{Literature Position and Contribution}

The literature on learning-based physical-layer communication has mainly followed two directions. The first direction studies neural networks as replacements or enhancements for classical communication blocks such as channel estimation, equalization, demapping, and detection \cite{oshea2017physical,simeone2018brief,wang2017wireless}. The second direction investigates end-to-end or receiver-side learning under more practical wireless assumptions, including over-the-air operation, software-defined radio platforms, and deep receiver structures \cite{dorner2018overair,honkala2021deeprx,restuccia2020physical,huang2020physical}.

In addition to these general neural receiver studies, FTN signaling has recently become an important test case for learning-based detection because it deliberately introduces controlled intersymbol interference. CNN-, GRU-, domain-aware CNN-, and attention-enhanced recurrent neural detector studies have shown that learning-based FTN receivers can exploit ISI-related sequence structures while addressing computational complexity \cite{tokluoglu2025standalone,tokluoglu2026gru,tokluoglu2025domainaware,tokluoglu2026bigru}. This line of work is closely related to the motivation of the present study because it connects neural detection performance with computational cost and practical receiver design.

These studies demonstrate the potential of deep learning in communication systems, but they also highlight an important practical issue: a neural receiver must be deployable. A receiver that performs well in offline simulation may still be unsuitable for real-time hardware if it has excessive memory demand, high inference latency, or large energy consumption. This issue is particularly important for embedded modems, SDR platforms, IoT nodes, FPGA implementations, and edge communication devices.

The contribution of this paper is positioned at this point. Instead of presenting AWGN-BPSK as a difficult channel model, this work uses it as a controlled and theoretically verifiable benchmark for parameter-efficiency analysis. Since the optimum detector is known, the experiment can focus on the architectural question: how many parameters are required for a neural receiver to reproduce a known decision behavior? This interpretation differs from a standard accuracy-only comparison. The proposed common-loss view evaluates whether a compact KAN receiver can reach a practical operating region with fewer parameters than an MLP baseline.

Therefore, the novelty of this work is not the AWGN-BPSK channel itself. The novelty is the receiver-design perspective: common-loss based comparison connects neural receiver accuracy with real-time implementation constraints. While many learning-based receiver studies emphasize BER or reconstruction error alone, this paper treats the parameter budget as a primary design variable. This provides a foundation for later extensions to fading, nonlinear, synchronization-impaired, or FTN channels where compact learnable receivers may become more practically valuable.

\section{Real-Time Receiver Motivation}

A possible criticism of neural receiver studies on AWGN-BPSK is that the problem is already solved. This criticism is valid only if the purpose is to replace the classical detector. In this work, AWGN-BPSK is not treated as a difficult final application. It is used as a theoretically verifiable calibration problem. If a neural receiver cannot reproduce the expected sign-like decision behavior in this controlled case, its reliability in more difficult channels would be questionable.

The stronger engineering question is not whether a neural network can outperform the optimum AWGN-BPSK detector. The relevant question is how compactly a neural receiver architecture can represent a known receiver behavior while reaching a comparable operating region. This shift is important for real-time communication systems. In a practical receiver, every symbol or sample must be processed within a strict timing budget. A model with fewer trainable parameters can reduce memory footprint, parameter access, inference latency, and energy consumption. It can also simplify implementation on FPGA block RAM, ASIC memory, DSP units, embedded processors, or SDR platforms.

Thus, parameter efficiency is not only a machine learning metric. It is a receiver design criterion. Once a target BER is reached, further reduction in regression loss may not justify a large increase in parameter count. This motivates the common-loss analysis used in this paper.

\section{System Model and Neural Receiver}

\subsection{BPSK over AWGN}

In BPSK modulation, binary information is mapped to antipodal symbols:
\begin{equation}
x \in \{-1,+1\}.
\end{equation}
The AWGN channel output is
\begin{equation}
y=x+n,
\end{equation}
where $n$ is a zero-mean Gaussian noise sample with variance $\sigma^2$. Under coherent reception and equal symbol probabilities, the optimum hard-decision detector is
\begin{equation}
\hat{x}=\mathrm{sign}(y).
\end{equation}
The theoretical BER is
\begin{equation}
P_b=Q\left(\sqrt{\frac{2E_b}{N_0}}\right),
\end{equation}
where $Q(\cdot)$ is the Gaussian Q-function, $E_b$ is bit energy, and $N_0$ is the noise power spectral density.

This expression provides the analytical reference. In this study, the learned receiver is expected to follow the same BER trend. This does not imply that the neural receiver is superior to the classical detector. Instead, it verifies whether the neural model learns a meaningful approximation of the correct receiver behavior.

\subsection{Neural Receiver Formulation}

The neural receiver is formulated as a supervised regression model. Given a received sample $y_i$, the receiver estimates the transmitted BPSK symbol:
\begin{equation}
\hat{x}_i=f_{\theta}(y_i).
\end{equation}
The model is trained using MSE:
\begin{equation}
\mathcal{L}_{\mathrm{MSE}}=\frac{1}{N}\sum_{i=1}^{N}(\hat{x}_i-x_i)^2.
\end{equation}
For BER evaluation, the continuous output is converted into a hard decision:
\begin{equation}
\tilde{x}_i=\begin{cases}
+1, & \hat{x}_i \geq 0,\\
-1, & \hat{x}_i < 0.
\end{cases}
\end{equation}
The BER is then computed as
\begin{equation}
\mathrm{BER}=\frac{1}{N}\sum_{i=1}^{N}\mathbb{I}\{\tilde{x}_i\neq x_i\}.
\end{equation}

This separation between regression loss and BER is important. MSE measures the continuous symbol estimation error, whereas BER measures final decision correctness. In hard-decision systems, two models may have similar BER even if their MSE values are different. Therefore, a model with slightly higher MSE can still be practical if it reaches the same decision region with fewer parameters.

\section{Dataset and Experimental Setup}

A synthetic Nyquist-rate BPSK dataset is generated for
\begin{equation}
\mathrm{SNR}\in\{7,8,9,10\}\ \mathrm{dB}.
\end{equation}
Random bits are mapped to BPSK symbols, AWGN is added, and the received samples are divided into training, validation, and test sets. Normalization statistics are computed only from the training set:
\begin{equation}
y_{\mathrm{norm}}=\frac{y-\mu_{\mathrm{train}}}{\sigma_{\mathrm{train}}}.
\end{equation}

Using only training-set statistics for normalization avoids information leakage from validation and test partitions. This is important because the purpose of the benchmark is to compare receiver architectures under identical and reproducible conditions. Each model therefore observes the same normalized input distribution and is evaluated using the same hard-decision rule.

\begin{figure}[!t]
    \centerline{\figmaybe{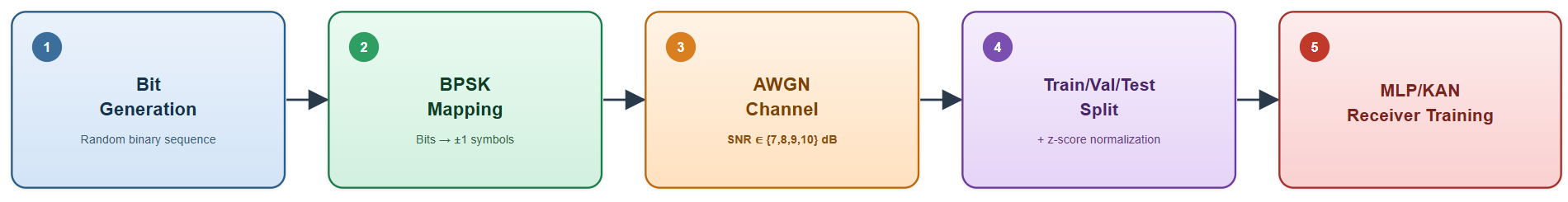}{Experimental pipeline placeholder.}}
    \caption{Overall experimental pipeline for the proposed neural receiver benchmark. Random bits are mapped to BPSK symbols, transmitted through an AWGN channel at multiple SNR levels, normalized using training-set statistics, split into training/validation/test subsets, and used for MLP and KAN receiver evaluation.}
    \label{fig:pipeline}
\end{figure}

\begin{table}[!t]
\caption{Dataset and Evaluation Configuration}
\label{tab:dataset}
\centering
\scriptsize
\setlength{\tabcolsep}{4pt}
\renewcommand{\arraystretch}{1.05}
\begin{tabular}{lc}
\toprule
\textbf{Property} & \textbf{Configuration} \\
\midrule
Modulation & BPSK $(\pm 1)$ \\
Channel & AWGN \\
Total samples & $4\times10^6$ \\
Train / Val / Test & $3.2{\times}10^6$ / $4{\times}10^5$ / $4{\times}10^5$ \\
SNR values & 7, 8, 9, 10 dB \\
Input & Noisy received sample \\
Target & Transmitted symbol \\
Learning task & Regression \\
Final decision & Sign threshold \\
Metrics & BER, MSE, parameter count \\
\bottomrule
\end{tabular}
\end{table}

The MLP receiver consists of dense layers and fixed nonlinear activation functions. It is represented as
\begin{equation}
\hat{x}=f_{\mathrm{MLP}}(y;\theta_{\mathrm{MLP}}).
\end{equation}
The MLP receiver represents the conventional dense neural baseline. Its approximation capacity can be increased by widening the hidden layers, but this also increases the number of scalar weights and biases. Therefore, the MLP baseline provides a useful reference for observing the trade-off between receiver accuracy and parameter budget.

The KAN receiver uses learnable univariate edge functions and is represented as
\begin{equation}
\hat{x}=f_{\mathrm{KAN}}(y;\theta_{\mathrm{KAN}}).
\end{equation}
The KAN receiver allocates trainable functions to edges rather than relying only on scalar edge weights and fixed node activations. This changes how nonlinear approximation capacity is distributed across the model. In this study, the KAN grid size controls the flexibility of the learnable univariate functions, while the hidden size controls the number of such functions.

All models are trained using the same data partitions, preprocessing, loss function, and BER evaluation rule. This ensures that the comparison focuses on architecture and parameter efficiency rather than differences in experimental protocol. The controlled experimental setup is intentional because it isolates the compact representation capability of the receiver architecture.

The use of a synthetic dataset also improves reproducibility because the modulation format, channel impairment, SNR range, target symbols, and decision rule are fully controlled. This prevents the comparison from being affected by dataset-specific labeling errors, uncontrolled channel variations, or implementation-dependent receiver assumptions. As a result, the observed differences between MLP and KAN receivers can be attributed mainly to the architectural representation and parameter allocation mechanisms rather than to external data variability.

\section{Common-Loss Parameter Efficiency}

The key methodological point of this work is common-loss comparison. Instead of selecting only the model with the minimum test loss, the analysis asks which model reaches a comparable practical performance region with fewer trainable parameters.

This criterion is particularly suitable for communication receivers because receiver design is usually constraint-driven. In practice, a receiver is first expected to satisfy a required BER region, and only then are implementation cost, latency, and energy consumption optimized. Therefore, comparing models only at their minimum achievable loss may obscure the more relevant engineering question: which architecture reaches the required operating point with the lowest implementation burden?

Another advantage of the common-loss formulation is that it separates architectural efficiency from the absolute difficulty of the benchmark. In the AWGN-BPSK case, the optimum decision rule is simple and analytically known; however, this simplicity makes the parameter requirement of each neural architecture more visible. If a model family requires a substantially larger parameter budget even for a known receiver mapping, this may indicate a less efficient representation for real-time deployment. Conversely, a compact model that reaches the same operating region suggests that its parameterization is well matched to the receiver function being approximated.

Let $\mathcal{F}$ denote a model family such as MLP or KAN. For a target loss level $\mathcal{L}^{*}$, the minimum parameter requirement can be expressed as
\begin{equation}
P_{\mathcal{F}}(\mathcal{L}^{*})=
\min_{\theta\in\mathcal{F}} P(\theta)
\quad
\mathrm{s.t.}
\quad
\mathcal{L}(\theta)\leq \mathcal{L}^{*},
\end{equation}
where $P(\theta)$ is the number of trainable parameters.

The relative parameter reduction between an MLP and a KAN receiver is
\begin{equation}
R_{\mathrm{param}}=\frac{P_{\mathrm{MLP}}-P_{\mathrm{KAN}}}{P_{\mathrm{MLP}}}\times100\%.
\end{equation}
For the strongest compact KAN case in this study, the MLP baseline contains 8513 trainable parameters, while the KAN receiver with $h=8$ and $g=24$ contains 3281 trainable parameters. The parameter reduction is therefore
\begin{equation}
R_{\mathrm{param}}=\frac{8513-3281}{8513}\times100\approx61.5\%.
\end{equation}

In communication receiver design, a target BER is often specified before implementation. Once this BER region is reached, the remaining design problem becomes minimizing complexity under the required performance constraint. The common-loss view follows this engineering logic by treating parameter count as a constrained receiver-design objective rather than as an auxiliary statistic.

A reduction of this magnitude is practically meaningful. It reduces the storage required for model coefficients and can reduce memory-access pressure during inference. In real-time receivers, memory access and arithmetic operations directly affect latency and energy consumption. Therefore, parameter reduction is not simply a numerical model-compression result; it changes the implementation profile of the receiver.

This is especially relevant for SDR, FPGA, ASIC, IoT, and embedded communication devices. In such systems, the receiver is not selected only by offline error performance. It must satisfy target BER under memory, latency, and power constraints. Therefore, a compact KAN receiver that reaches the same practical BER region as a larger MLP can be preferable even if the MLP has slightly lower MSE.

\section{Results and Discussion}

\subsection{BER Performance over SNR}

\begin{figure}[!t]
    \centerline{\figmaybe{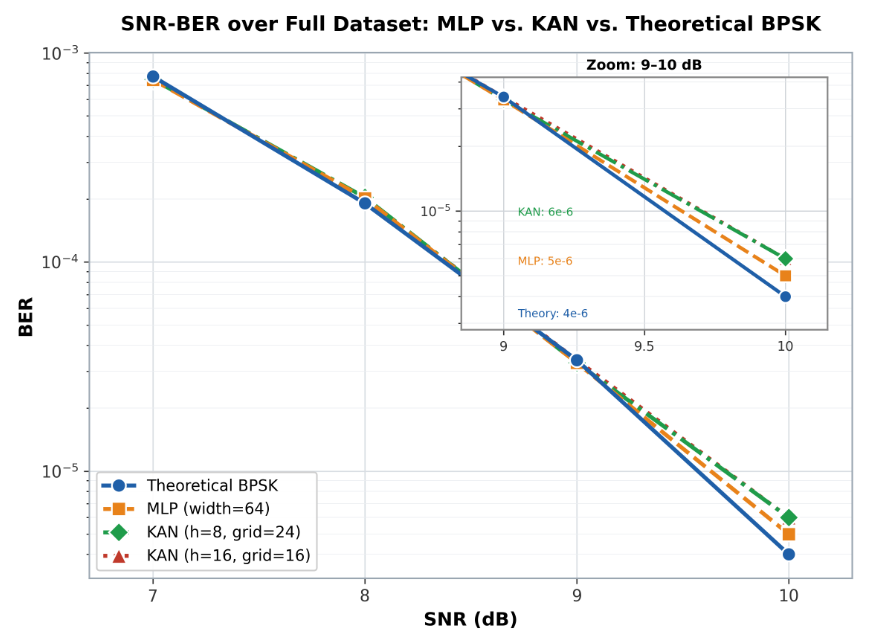}{BER versus SNR placeholder.}}
    \caption{BER versus SNR comparison of the theoretical BPSK reference, MLP receiver, and representative KAN receiver configurations over the full test dataset. The learned receivers closely follow the theoretical AWGN-BPSK trend, supporting the validity of the neural receiver benchmark.}
    \label{fig:ber}
\end{figure}

Fig.~\ref{fig:ber} shows the BER performance over SNR. Both MLP and KAN receivers follow the expected AWGN-BPSK trend: BER decreases as SNR increases. This confirms that both neural receiver families reproduce the known detection behavior.

The purpose of this result is not to claim that a neural receiver is required for AWGN-BPSK. Rather, the purpose is to verify that the learned models approximate the known sign-like receiver mapping. Once this behavior is verified, the more important question becomes how many parameters are required to reach the practical BER region.

\begin{table}[!t]
\caption{BER Results over SNR}
\label{tab:ber}
\centering
\scriptsize
\renewcommand{\arraystretch}{1.08}
\resizebox{\columnwidth}{!}{%
\begin{tabular}{lccccc}
\toprule
\textbf{Receiver} & \textbf{Params} & \textbf{7 dB} & \textbf{8 dB} & \textbf{9 dB} & \textbf{10 dB} \\
\midrule
MLP, $w=64$       & 8513 & $7.46\mathrm{e}{-4}$ & $2.02\mathrm{e}{-4}$ & $3.30\mathrm{e}{-5}$ & $5.00\mathrm{e}{-6}$ \\
KAN, $h=8,g=24$  & 3281 & $7.49\mathrm{e}{-4}$ & $2.04\mathrm{e}{-4}$ & $3.30\mathrm{e}{-5}$ & $6.00\mathrm{e}{-6}$ \\
KAN, $h=16,g=16$ & 8313 & $7.49\mathrm{e}{-4}$ & $2.01\mathrm{e}{-4}$ & $3.40\mathrm{e}{-5}$ & $6.00\mathrm{e}{-6}$ \\
\bottomrule
\end{tabular}%
}
\end{table}

Table~\ref{tab:ber} shows that the learned receivers operate in the same BER region over the considered SNR range. The compact KAN model with 3281 parameters remains very close to the MLP baseline with 8513 parameters. This observation supports the common-loss interpretation: the focus is not only which model gives the smallest numerical BER at each SNR, but which model reaches a comparable operating region with fewer trainable parameters.

\subsection{SNR-wise Interpretation}

The SNR-wise results show that the learned receivers follow the expected AWGN-BPSK behavior. As the SNR increases from 7 dB to 10 dB, the BER values decrease consistently for both MLP and KAN receivers. This confirms that the neural models learn the same sign-like decision behavior as the classical hard-decision receiver. Small numerical deviations from the theoretical BER are expected because empirical BER is computed from a finite test set.

The important observation is that the compact KAN receiver remains in the same practical BER region as the MLP baseline across all SNR values, despite using substantially fewer trainable parameters. This SNR-wise consistency is important because it shows that the compact KAN receiver does not match the MLP baseline only at a single aggregate test point, but remains close across the evaluated SNR range. Therefore, the SNR-wise results support the common-loss interpretation rather than an accuracy-only comparison.

\subsection{Global Parameter-Efficiency Comparison}

The MSE provides additional information about continuous symbol estimation quality by measuring how close the receiver output is to the transmitted BPSK symbol. However, BER depends on whether the output remains on the correct side of the decision threshold. After verifying the SNR-dependent BER behavior, the next question is not whether the learned receivers can solve the benchmark, but how efficiently they do so. Therefore, Table~\ref{tab:metrics} summarizes the global parameter-performance trade-off.

\begin{table}[!t]
\caption{Global Comparison of Selected Receivers}
\label{tab:metrics}
\centering
\scriptsize
\setlength{\tabcolsep}{3.2pt}
\renewcommand{\arraystretch}{1.08}
\begin{tabular}{lccccc}
\toprule
\textbf{Family} & \textbf{W/H} & \textbf{Grid} & \textbf{Params} & \textbf{MSE} & \textbf{BER} \\
\midrule
MLP & 64 & -- & 8513  & $7.04{\times}10^{-4}$ & $2.50{\times}10^{-4}$ \\
KAN & 8  & 24 & 3281  & $6.98{\times}10^{-4}$ & $2.45{\times}10^{-4}$ \\
KAN & 16 & 16 & 8313  & $7.00{\times}10^{-4}$ & $2.47{\times}10^{-4}$ \\
KAN & 32 & 2  & 13771 & $7.22{\times}10^{-4}$ & $2.55{\times}10^{-4}$ \\
\bottomrule
\end{tabular}
\end{table}

Table~\ref{tab:metrics} is the central quantitative result of this study and provides the main evidence for the proposed parameter-efficiency argument. The compact KAN receiver with $h=8$ and $g=24$ achieves a test BER of $2.45\times10^{-4}$ using 3281 trainable parameters, whereas the MLP baseline achieves a test BER of $2.50\times10^{-4}$ using 8513 trainable parameters. Therefore, the compact KAN configuration reaches a comparable BER operating region with approximately 61.5\% fewer trainable parameters.

It is also important to evaluate MSE and BER together because they reflect different aspects of receiver behavior. MSE measures the continuous-valued symbol reconstruction quality, whereas BER measures the final binary decision after thresholding. For BPSK detection, a receiver output may have a slightly different continuous amplitude but still remain on the correct side of the decision boundary. Therefore, a compact model that preserves the final decision behavior can be practically useful even when small differences in regression loss are observed.

This result should be interpreted from a common-loss perspective rather than an accuracy-only perspective. The compact KAN model is not presented as the universally best architecture across all possible configurations; rather, it is the smallest verified configuration in this study that reaches the same practical BER region as the selected MLP baseline. Therefore, the parameter-efficiency claim is based on jointly considering test BER, test MSE, and trainable parameter count.

In real-time communication receivers, this difference is practically meaningful. A lower parameter count reduces the memory required to store the receiver model and can reduce the number of parameter accesses during inference. Since memory access, data movement, and arithmetic operations directly affect latency and energy consumption, reaching the same BER region with fewer parameters can improve the feasibility of neural receivers on embedded processors, SDR platforms, FPGA/ASIC implementations, IoT devices, and edge communication systems.

\subsection{Real-Time Interpretation}

The main advantage of KAN over the selected MLP baseline is compact representation at a comparable receiver operating point rather than universal numerical superiority. As shown in Table~\ref{tab:metrics}, the compact KAN model reaches nearly the same BER region as the MLP baseline while using substantially fewer trainable parameters.

This result is relevant for real-time receivers because parameter count affects stored coefficient memory, parameter access, computational burden, latency, and energy consumption. Once a target BER region is reached, a small improvement in offline MSE may not justify a much larger model. Under this common-loss view, the compact KAN receiver is attractive because it reaches the same useful communication-performance region with a smaller implementation cost.

\subsection{What the Results Contribute}

The results contribute to the literature by showing that a theoretically known receiver problem can be used as a controlled benchmark for validating neural receiver behavior and comparing architectures under implementation-aware constraints. Rather than selecting a receiver only by the lowest offline MSE, the proposed common-loss view jointly considers BER, MSE, and parameter count.

This perspective is relevant for future extensions to sequence equalization, FTN signaling, OFDM demapping, MIMO detection, nonlinear channels, and hardware-impaired reception. Existing FTN neural detector studies already show that CNN-, GRU-, domain-aware CNN-, and attention-enhanced recurrent receivers can address controlled ISI while considering computational efficiency \cite{tokluoglu2025standalone,tokluoglu2026gru,tokluoglu2025domainaware,tokluoglu2026bigru}. In such cases, compact nonlinear KAN representations may offer stronger practical benefit.

\section{Limitations and Future Work}

This study has intentional limitations. The channel is AWGN-only, the modulation is BPSK, and the receiver input is one-dimensional. The system does not include fading, intersymbol interference, synchronization mismatch, carrier-frequency offset, phase noise, nonlinear distortion, quantization, or hardware impairments. In addition, the classical sign detector is already optimal for this channel.

These limitations should be interpreted as controlled design choices rather than omissions. The AWGN-BPSK setting is used to establish a reproducible parameter-efficiency evaluation protocol under a known theoretical reference before introducing additional channel impairments.

Future work will extend the same common-loss methodology to more challenging scenarios such as FTN signaling, multipath fading, nonlinear channels, OFDM demapping, MIMO detection, and synchronization-impaired reception. FTN signaling is particularly relevant because it introduces controlled intersymbol interference, where existing CNN-, GRU-, domain-aware CNN-, and attention-enhanced recurrent detectors have already shown the value of learning-based receiver design \cite{tokluoglu2025standalone,tokluoglu2026gru,tokluoglu2025domainaware,tokluoglu2026bigru}. In such scenarios, compact KAN-based receivers may provide stronger practical benefit.

\section{Conclusion}

This paper presented a common-loss based parameter-efficient neural receiver analysis for Nyquist-rate BPSK reception over AWGN channels. The AWGN-BPSK scenario was intentionally selected as a controlled benchmark because the optimum hard-decision detector and theoretical BER are known. Therefore, the purpose was not to replace the classical detector, but to evaluate how compactly neural receiver architectures can reproduce a physically meaningful decision behavior.

The results show that both MLP and KAN receivers follow the expected AWGN-BPSK BER trend. The compact KAN configuration with $h=8$ and $g=24$ reaches a comparable BER operating region using 3281 trainable parameters, whereas the MLP baseline uses 8513 parameters. This corresponds to approximately 61.5\% fewer trainable parameters at a comparable BER operating point.

The main conclusion is that common-loss analysis provides a useful way to compare neural receivers under implementation constraints. In real-time systems, offline MSE alone is not sufficient; memory footprint, parameter access, latency, energy consumption, and hardware complexity also matter. Future work will extend this methodology to more challenging communication scenarios, particularly FTN signaling, where controlled ISI may make compact nonlinear receiver representations more valuable.



\begin{thebibliography}{00}

\bibitem{proakis2007digital}
J. G. Proakis and M. Salehi,
\emph{Digital Communications},
5th ed. New York, NY, USA: McGraw-Hill, 2007.

\bibitem{sklar2001digital}
B. Sklar,
\emph{Digital Communications: Fundamentals and Applications},
2nd ed. Upper Saddle River, NJ, USA: Prentice Hall PTR, 2001.

\bibitem{oshea2017physical}
T. J. O'Shea and J. Hoydis,
``An Introduction to Deep Learning for the Physical Layer,''
\emph{IEEE Transactions on Cognitive Communications and Networking},
vol. 3, no. 4, pp. 563--575, Dec. 2017,
doi: 10.1109/TCCN.2017.2758370.

\bibitem{dorner2018overair}
S. D\"orner, S. Cammerer, J. Hoydis, and S. ten Brink,
``Deep Learning Based Communication Over the Air,''
\emph{IEEE Journal of Selected Topics in Signal Processing},
vol. 12, no. 1, pp. 132--143, Feb. 2018,
doi: 10.1109/JSTSP.2017.2784180.

\bibitem{honkala2021deeprx}
M. Honkala, D. Korpi, and J. M. J. Huttunen,
``DeepRx: Fully Convolutional Deep Learning Receiver,''
\emph{IEEE Transactions on Wireless Communications},
vol. 20, no. 6, pp. 3925--3940, June 2021,
doi: 10.1109/TWC.2021.3054520.

\bibitem{simeone2018brief}
O. Simeone,
``A Very Brief Introduction to Machine Learning With Applications to Communication Systems,''
\emph{IEEE Transactions on Cognitive Communications and Networking},
vol. 4, no. 4, pp. 648--664, Dec. 2018,
doi: 10.1109/TCCN.2018.2881442.

\bibitem{wang2017wireless}
T. Wang, C.-K. Wen, H. Wang, F. Gao, T. Jiang, and S. Jin,
``Deep Learning for Wireless Physical Layer: Opportunities and Challenges,''
\emph{China Communications},
vol. 14, no. 11, pp. 92--111, Nov. 2017,
doi: 10.1109/CC.2017.8233654.

\bibitem{oshea2018radio}
T. J. O'Shea, T. Roy, and T. C. Clancy,
``Over-the-Air Deep Learning Based Radio Signal Classification,''
\emph{IEEE Journal of Selected Topics in Signal Processing},
vol. 12, no. 1, pp. 168--179, Feb. 2018,
doi: 10.1109/JSTSP.2018.2797022.

\bibitem{restuccia2020physical}
F. Restuccia and T. Melodia,
``Deep Learning at the Physical Layer: System Challenges and Applications to 5G and Beyond,''
\emph{IEEE Communications Magazine},
vol. 58, no. 10, pp. 58--64, Oct. 2020,
doi: 10.1109/MCOM.001.2000107.

\bibitem{huang2020physical}
H. Huang, S. Guo, G. Gui, Z. Yang, J. Zhang, H. Sari, and F. Adachi,
``Deep Learning for Physical-Layer 5G Wireless Techniques: Opportunities, Challenges and Solutions,''
\emph{IEEE Wireless Communications},
vol. 27, no. 1, pp. 214--222, Feb. 2020,
doi: 10.1109/MWC.2019.1900027.

\bibitem{tokluoglu2025standalone}
O. Tokluoglu, E. Cavus, E. Bedeer, and H. Yanikomeroglu,
``A Novel CNN-Based Standalone Detector for Faster-Than-Nyquist Signaling,''
\emph{IEEE Transactions on Communications},
vol. 73, no. 12, pp. 14316--14331, Dec. 2025,
doi: 10.1109/TCOMM.2025.3602366.

\bibitem{tokluoglu2026gru}
O. Tokluoglu, A. Cicek, E. Cavus, E. Bedeer, and H. Yanikomeroglu,
``GRU-Based Sequence Detection for Faster-than-Nyquist Signaling,''
\emph{IEEE Open Journal of Vehicular Technology},
vol. 7, pp. 565--581, Jan. 14, 2026,
doi: 10.1109/OJVT.2026.3653504.

\bibitem{tokluoglu2025domainaware}
O. Tokluoglu, E. Cavus, E. Bedeer, and H. Yanikomeroglu,
``A Novel Domain-Aware CNN Architecture for Faster-than-Nyquist Signaling Detection,''
in \emph{Proc. IEEE 36th International Symposium on Personal, Indoor and Mobile Radio Communications (PIMRC)},
Istanbul, Turkiye, 2025, pp. 1--6,
doi: 10.1109/PIMRC62392.2025.11274560.

\bibitem{tokluoglu2026bigru}
O. Tokluoglu, E. Cavus, E. Bedeer, and H. Yanikomeroglu,
``A Novel Position-Aware Attention-Enhanced Bi-GRU Detector for Faster-than-Nyquist Signaling,''
\emph{IEEE Communications Letters},
Early Access, 2026,
doi: 10.1109/LCOMM.2026.3704639.

\bibitem{cybenko1989approximation}
G. Cybenko,
``Approximation by superpositions of a sigmoidal function,''
\emph{Mathematics of Control, Signals, and Systems},
vol. 2, no. 4, pp. 303--314, 1989,
doi: 10.1007/BF02551274.

\bibitem{liu2025kan}
Z. Liu, Y. Wang, S. Vaidya, F. Ruehle, J. Halverson, M. Solja\v{c}i\'c, T. Y. Hou, and M. Tegmark,
``KAN: Kolmogorov--Arnold Networks,''
in \emph{Proc. International Conference on Learning Representations (ICLR)}, 2025.

\bibitem{kolmogorov1957representation}
A. N. Kolmogorov,
``On the representation of continuous functions of many variables by superposition of continuous functions of one variable and addition,''
\emph{Doklady Akademii Nauk SSSR},
vol. 114, pp. 953--956, 1957.

\bibitem{arnold1957functions}
V. I. Arnold,
``On functions of three variables,''
\emph{Doklady Akademii Nauk SSSR},
vol. 114, no. 4, pp. 679--681, 1957.

\bibitem{deboor1978splines}
C. de Boor,
\emph{A Practical Guide to Splines}.
New York, NY, USA: Springer, 1978.

\end{thebibliography}
\end{document}